\documentclass[preprint]{aastex}
\input epsf

\shorttitle{Globular Clusters in the Fornax Dwarf}
\shortauthors{Strader et al.}

\begin{document}

\title{Spectroscopy of Globular Clusters in the Fornax dwarf galaxy}

\author{Jay Strader and Jean P. Brodie}
\affil{UCO/Lick Observatory, University of California, Santa Cruz, CA 95064}
\email{strader@ucolick.org, brodie@ucolick.org}

\author{Duncan A. Forbes and Michael A. Beasley}
\affil{Centre for Astrophysics and Supercomputing, Swinburne University, Hawthorn VIC 3122, Australia}
\email{dforbes@astro.swin.edu.au, mbeasley@astro.swin.edu.au}
\and
\author{John P. Huchra}
\affil{Harvard-Smithsonian Center for Astrophysics, Cambridge, MA 02138}
\email{huchra@cfa.harvard.edu}

\begin{abstract}

We present low resolution, integrated Keck spectra of the five globular clusters (GCs) of the Fornax dwarf spheroidal galaxy. We find a tentative age
spread among the clusters, with the GC H5 younger by 2--3 Gyr than the others. The clusters generally appear to be very metal-poor ([Fe/H] $\sim$ $-$1.8),
with the cluster H4 slightly more metal-rich at [Fe/H] = $-$1.5. We speculate that cluster H4 is similar to the Galactic GC Ruprecht 106, which lacks the
[$\alpha$/Fe] enhancement typical among metal-poor GCs in the Milky Way. High-resolution spectroscopy of individual cluster and field stars will be needed to
sort out the surprisingly complex history of GC formation and evolution in this galaxy.

\end{abstract}


\section{Introduction}

The Fornax dwarf spheroidal (dSph) galaxy has one of the most complex star formation histories of any of the Local Group dSphs \citep{S98}. While a
few of these galaxies seem to have only old stars (e.g., Draco, Sculptor), such dSphs as Leo I and Carina show star formation up to $\sim 2$ Gyr ago
\citep{D02}. In contrast, several studies (Stetson et al.~1998; Buonanno et al.~1999, hereafter B99) have found an age range of $\sim$ 0.5--12 Gyr
in the Fornax field, dominated by intermediate-age stars. This complex history could be due in part to the galaxy's high luminosity (and mass); at M$_{V} \sim
-13.5$ it is second only to the Sagittarius dSph.

Fornax is unique among the Local Group dSphs in that it has a significant population of globular clusters---five---which results in a specific frequency
$S_{N} \sim 20$. This is the highest $S_{N}$ known for any galaxy. However, increasing evidence suggests that other dSphs may have once had their own GCs
as well. Sagittarius, for example, is now known to have at least four GCs (M54, Ter 4, Ter 7, Arp 2). \citet{D00} and \citet{M02} have argued that the
tidally  disrupting GC Pal 12 may be associated with Sagittarius as well. More speculatively, \citet{YL02} have proposed that a subset of the Oosterhoof II
Milky Way GCs (including M92 and M68) may have been tidally stripped from the Large Magellanic Cloud. This suggests that GC populations in dSph/dIrr
galaxies may have been more common is the past, but that Fornax was ``dynamically lucky'' in that its distance from the Galactic Center ($\sim 140$ kpc) may
have prevented similar stripping of its GCs. Thus a close study of the Fornax GC system (GCS) could shed light on what may have been a common phenomenon
among dwarf galaxies. While the discovery of other such systems would not explain the large difference in $S_{N}$ between Fornax and, for example, regular
dwarf spheroidals, it would help address why Fornax currently appears to be a singular object.

For its small size, the Fornax GCS is surprisingly complex. For example, \citet{RR94} found that while the radial profiles of clusters 1 and 2 (hereafter we
refer to the GCs as H1--H5, according to Hodge 1961) are well-fit by King profiles with tidal truncation (though the H1 sample was poor),
they could not find any evidence of such truncation in H3--H5. While it is difficult to distinguish between a large tidal radius and no truncation
at all, even a large difference in tidal radii between the two groups is not easily explained. There appear to be no simple correlations between these
properties and galactrocentric distance, metallicity, or age. The radial profile of H5 is even more complex: Mackey \& Gilmore (2003) find that the core
of the cluster is best fit by a power law (rather than the flat, constant density core of a King model) and suggest that H5 may be a post core-collapse
GC.

\citet{O00} show that the expected timescale for mass segregation in the galactic potential is $\sim 1$ Gyr and that the GCs should have sunk to the
center of the galaxy. That this is not the case is evidence that some other mechanism is kinematically exciting the clusters into their present orbit.
They suggest that the most likely scenario is one in which tidal effects from the Galaxy torque the cluster orbits. Though dynamical friction within
Fornax will counteract this force, the present widely spread GC locations would be possible only if the galaxy were undergoing significant mass loss (at
least 30\% of the original total). This might indicate that a Hubble time ago Fornax was much more massive, possibly explaining its high GC specific
frequency. Were this scenario true, we would expect to see tidally stripped stars with high proper motions (and chemical similarities to Fornax field stars)
in the area surrounding the galaxy. No such studies have yet been published. A recent proper motion study of Fornax \citep{P02} suggests that it may not
even be bound to the Galaxy, although a bounded low-eccentricity or even circular orbit is consistent with the data if the galaxy is near perigalacticon.
The Keplerian period for such an orbit would be 4--8 Gyr; the hypothetical mass loss could have peaked during the last perigalactic passage.

In this paper, we present a spectroscopic survey of the Fornax GCs, deriving ages and metallicities, with a view to constraining the
formation and evolution of the galaxy.

\section{Observations and Data Reduction}

The observations were made using the Low Resolution Imaging Spectrometer \citep{O95} in longslit mode on the Keck I telescope. All five clusters were observed
during a two night run in December 1996. A single 300 s exposure was taken for each of the clusters. These observations were supplemental to our main program.
The $V$ band magnitudes of the five GCs are H1 (15.6), H2 (13.5), H3 (12.6), H4 (13.4), and H5 (13.4), all from Webbink (1985).

All observations were made with 600 line mm$^{-1}$ gratings, blazed at 5000 \AA. This gave a spectral resolution of $\sim$ 6 \AA\ and a dispersion of
1.28 \AA\ pixel$^{-1}$. The total wavelength range was generally 3800--6200 \AA, but the usable range was often shorter by several hundred
angstroms, due both to low signal-to-noise (S/N) ratios for several objects as well as the poor instrumental response of the LRIS ``red arm'' at
the blue end of the spectrum (these spectra were taken before installation of the ``blue arm'').

The standard data reduction was performed using IRAF\footnote{IRAF is distributed by the National Optical Astronomy Observatories, which are operated by the
Association of Universities for Research in Astronomy, Inc., under cooperative agreement with the National Science Foundation.}. Raw images were debiased and then
flatfielded, using a normalized composite sky flat. After cosmic ray removal using the IRAF task \emph{cosmicrays}, the spectra were wavelength calibrated with HgKr
comparison arc lamp spectra. These spectra were flux-calibrated using the flux standard Feige 24. The spectra were corrected to zero redshift before the measurement of
any line indices. For reference, Figure 1 shows the spectra of the five GCs.

\section{Lick/IDS Indices}

Elemental abundances were measured using Lick/IDS absorption-line indices (Trager et al.~1998 and references therein), from spectra convolved with
wavelength-dependent Gaussian kernel to more closely match the Lick/IDS resolution. See \citet{LB02} and \citet{S03} for more extensive discussions of
this issue.

We first measured a set of Lick/IDS indices (listed in Table 1) for comparison to stellar evolutionary models. Though we did measure indices for the faint
cluster H1, the errors are too large for their inclusion in the tables, and we exclude H1 from subsequent analysis. To check that the indices were
self-consistent, we plotted the major metallicity indicators (Mg$_{2}$, Mgb, Fe5335) against each other. In each case, within the error bars, the indices fell
along a 1:1 relation.

Our measurements of Lick/IDS Mg and Ca features have large index errors, translating into 0.3--0.4 dex errors in [$\alpha$/Fe], making it impossible to determine
whether a given cluster is truly $\alpha$-enhanced. An additional complicating factor is that $\alpha$-enhanced SSP models (e.g., Milone et al.~2000; Thomas et
al.~2002) converge at very low metallicities. For these reasons we were unable to make any accurate estimates of [$\alpha$/Fe] for the Fornax GCs.

\section{Metallicities}

\citet[hereafter BH90]{BH90} defined a procedure for estimating the metallicity of globular clusters with low resolution spectra by taking a weighted mean
of various elemental absorption-line indices sensitive to [Fe/H]. Indices defined over a wavelength range of interest are calculated with respect to a
pseudocontinuum, defined using regions to either side of the feature bandpass.

No resolution correction was made for measuring BH90 metallicity indices, since the BH90 calibration primarily used data near our resolution or
better. Then, as described in \S 2, the spectra were flux calibrated before any abundance measurements were made. The minor role of flux 
calibration in measuring indices has been extensively documented \citep{F85, KP98, LB02}. To verify this, we measured several strong features on
the spectra both before and after flux calibration. The differences between the fluxed and unfluxed line indices were generally $\sim$ 3--5\% of
the error bars, and thus insignificant.

We estimated [Fe/H] for clusters H2--H5 according to the formulae in BH90, with the exception of a slight modification in calculating the
variance, as described in \citet{LB02}. The feature indices used in the spectroscopic measurements are listed in Table 2. The resulting individual
index [Fe/H] estimates, as well as the composite values, are listed in Table 3.

\section{Ages}

To estimate the ages of the GCs in our sample, we plotted various combinations of the age-sensitive indices (e.g H$\beta$ and H$\gamma$) against the metallicity
features (e.g., Mg$_{2}$, Mgb, Fe5270, Fe5335). Figure 2 shows H$\beta$ plotted against a combined Fe5270-Fe5335 index (denoted $<$Fe$>$) for clusters H2--H5.
Superimposed on the plots are models obtained from Thomas et al.~(2002). The GCs H2--H4 are coeval within the errors, while H5 appears to be younger by a couple of Gyr
at the 2$\sigma$ level. An H$\beta$ vs [MgFe] diagram (Figure 3) supports the slightly younger age of H5. One advantage to using the composite index [MgFe] is that is
it insensitive to [$\alpha$/Fe] variations. The lack of \emph{absolute} agreement between the index measurements and Simple Stellar Population (SSP) models may be due
to overestimation of the rise in H$\beta$ absorption at very low metallicities (R. Schiavon, private communication). These diagrams provide support at the $\sim
2\sigma$ level for an age spread in the Fornax GCs, with H5 several Gyr younger than the other clusters, despite its relatively low metallicity.

Our Fornax data are compared to the Lick-corrected Large Magellanic Cloud (LMC) sample of globular clusters from Beasley, Hoyle \& Sharples (2002) on the Thomas et
al.~(2002) models in Figures 2 \& 3. In the H$\beta$-$<$Fe$>$ planes the Fornax clusters span similar ranges in metallicity, although in the mean the LMC sample is
slightly more metal-rich. Two of the LMC clusters show strong H$\beta$ absorption on the SSP grids; Beasley et al.~attribute this to blue horizontal branch (HB)
effects rather than younger ages. The Fornax cluster H5 also exhibits a slightly elevated H$\beta$ line-strength, with a very similar metallicity (and presumably age)
to the LMC cluster NGC~1916. The Buonanno et al. (1998) CMD of H5 clearly shows a blue HB. In principle, the stronger H$\beta$ index, and hence younger inferred age
of the cluster, could be due to the presence of the blue HB (e.g., Lee et al. 2000). However, we note that H2 and H3 have very similar blue HB morphologies and yet
have relatively weak H$\beta$ lines. Our Fornax data is not well-constrained by the SSP models in Figure 3, since three of the four clusters lie to the left/below the
grids. However, our general conclusion is that these GCs are old, with cluster H4 somewhat more metal-rich than the rest.

To probe these results, we undertook a similar exercise using the age-sensitive indices H$\gamma$ and H$\delta$, with inconclusive results. These indices are
can be less dependent on metallicity, if defined properly (for example, see the H$\gamma$ definition in Vazdekis \& Arimoto 1999). However, this reduced  
metal-sensitivity is only possible because the index range is very small (less than one Angstrom). This, in turn, means that the spectra must be very high S/N
to obtain an accurate index value. Another complicating factor is that there is a significant age-metallicity degeneracy in current SSP models for these indices at
very low [Fe/H] values, and thus the measured H$\gamma$ and H$\delta$ indices for our clusters are compatible with isochrones from 7--14 Gyr. The indices
H$\gamma_{A}$ and H$\delta_{A}$, as defined in Worthey \& Ottaviani (1997), are listed in Table 4.

For GCs H2-H5 we have also derived a metallicity estimate using the H$\beta$ vs $<$Fe$>$ model grid. In Table 5 we list this estimate (called the ``Lick
Indices'' method) along with that derived from the BH90 method (see \S 4).

\section{Comparison with Previous Works}

One immediately noticeable feature in Table 3 is how little spread there is in our metallicity estimates, amounting to only $\sim 0.3$ dex among clusters
H2--H5. This is in contrast to the larger spread ($\sim$ 0.6--0.9 dex) found by some early investigators (e.g., Dubath et al.~1992), which was primarily
driven by the high metallicity of H4 with respect to the other clusters. Compared to the Dubath et al.~[Fe/H] estimates, we find that H3 and H5 are
slightly more metal-rich, with [Fe/H] = $-$1.8 and $-$1.7 respectively, while H4 is somewhat more metal-poor at [Fe/H] = $-$1.5. The sum of these small
shifts substantially reduces the overall dispersion. Were we able to include H1, it would likely not affect this result, since previous investigations
uniformly have found that within the error bars H1 and H5 have the same metallicity \citep{D92, B85, B98}. For comparison purposes, we have listed our
[Fe/H] estimates together with past measurements using a variety of methods in Table 5. Most previous researchers find H4 to be slightly more metal-rich
that the other four GCs. The case of H4 is discussed further in \S 6.1. 

Several of the less common methods in Table 5 merit explanation: Dubath et al.~(1992) cross-correlated the cluster spectrum with that of a template;
calculating the area under the peak of the cross-correlation function is a measure of the effective temperature of the RGB of the cluster and hence of the
metallicity. Demers et al.~(1990) converted the $B - V$ color of H1 to a metallicity using the Zinn \& West (1984) relation. Finally, the ``Zinn Q''
method of Zinn \& Persson (1981) uses a transformation between the cluster metallicity and the quantity $Q_{39}$, which is a composite photometric measure
of the curvature and line blanketing in the cluster spectrum.

In terms of GC ages, we found the GCs to be uniformly old with H5 perhaps a few Gyr younger. The primary alternative sources of Fornax GC age measurements
are the color-magnitude diagrams (CMDs), the most accurate of which are constructed from HST images (B99, Buonanno et al.~1998). To compare our age
measurements with the CMD methods, we chose the cluster pair H3 and H5, as the error bars on H2 are somewhat larger and unknown parameters may be
affecting the H4 CMD. \citet{B98} found essentially no age difference between the two clusters using either the $\Delta V^{\rm{TO}}_{\rm{HB}}$ method
($0.1\pm2.8$ Gyr) or the $\Delta(V-I)^{\rm{TO}}_{\rm{RGB}}$ method ($0.1\pm0.8$ Gyr). However, the error bars on the measurements include \emph{only}
errors in the measurement of $V_{\rm{HB}}$ and $V_{\rm{TO}}$. If other sources of error are included, such as the uncertainty in the HB M$_{V}$-[Fe/H]
relation, the possible age spread of 2--3 Gyr found by this work agrees with the CMD findings within the error bars.

\subsection{The Strange Case of H4}

Early spectroscopic and photometric studies of H4 indicated it was relatively metal-rich at [Fe/H] $\sim -1.4$, compared to the other Fornax GCs, which
seemed to be uniformly metal-poor \citep{HC77, ZP81, D92}. Interestingly, this value is close the mean of the Fornax field ([Fe/H] = $-$1.36; Buonanno et
al.~1999). Detailed studies of the CMD of the cluster were difficult to make due to its high density and location near the center of the galaxy.
Ground-based attempts failed to achieve a CMD accurate enough to substantially constrain the metallicity, although CMD metallicity estimates were lower
than the spectroscopic values \citep{B85, B95}. However, B99 were able to construct a relatively clean CMD using HST Wide Field and Planetary Camera 2
(WFPC2) images. By comparing the slope of the cluster's RGB to that of Galactic GCs, they found the metallicity of H4 was [Fe/H] = $-2.01\pm0.20$, in
sharp disagreement with earlier measurements and our work.

B99 considered several explanations for the discrepancy. The first was a significantly larger reddening for H4 than for the other
clusters, but by comparing the H4 CMD to that of H2, they argued that the amount of reddening needed to account for the apparent metallicity
difference in the two methods was improbably high. Field star contamination was also considered as a possible factor, since the surrounding field 
stars are of a higher metallicity than the cluster stars (as ascertained from the HST CMD). This might especially be an issue for H4 as it lies
close to the center for the galaxy.

Their third explanation concerns the nature of the relationship between [Fe/H] and the RGB. Noting that the correlations between broadband
colors or elemental line indices and the metallicity of the cluster was dependent primarily on the nature of the RGB, and that the calibration of these
relations is based upon Galactic GCs, they argue that were H4 somehow different from the bulk of Galactic GCs, such relationships might not hold.
Thus, the spectroscopic [Fe/H] estimates might be invalid. To support this view, they show that the shape of the H4 CMD, with an extremely red HB and a
relatively bright SGB, differs substantially from that of the other Fornax GCs as well as that of a ``classic'' metal-poor Milky Way GC, M68, in that the
CMDs cannot be shifted to fit each other with reasonable metallicity and age differences. B99 suggest that an anomalous HB and SGB are the
source of the discrepancy in that they cause the integrated cluster spectrum to appear to be more metal-rich. While a parameter which \emph{only} affected
the HB and SGB and not the slope of the RGB could plausibly exist, B99 offer no suggestions.

B99 describe their fourth explanation of the difference as ``speculative.'' They note that with the appropriate reddening corrections, the CMD
of H4 is a close fit to that of Ruprecht 106 (hereafter R106), an outer halo Milky Way GC believed to be 3-4 Gyr younger than the typical old,
metal-poor Galactic GC. Indeed, there has been a similar discrepancy between the spectroscopic and RGB slope [Fe/H] measurements for this cluster,
with the former method yielding a much higher metallicity than the latter \citep{B97, B93}

Importantly, \citet{B97} also found that R106 had anomalous values of [O/Fe] and [$\alpha$/Fe]. The former measurement yielded a value of [O/Fe] =
$0.0\pm0.13$. In a typical metal-poor cluster [O/Fe] $\sim$ 0.3--0.4, although there is a great deal of variation both from cluster to cluster and
within some individual GCs \citep{C01}. A similar difference is seen in the $\alpha$-element abundance---\citet{B97} found [$\alpha$/Fe] = 
$-0.05\pm0.05$, compared to a typical value of [$\alpha$/Fe] $\sim 0.3$ \citep{C96}. \citet{B97} also found an $\alpha$-element underabundance in
the young outer-halo Galactic GC Pal 12, the only other Milky Way cluster for which a lack of $\alpha$-element enhancement is seen \citep{C01}.
No complete, deep $VI$ CMD of Pal 12 currently exists (the closest is Rosenberg et al. 1998).

These results suggest a detailed examination of [$\alpha$/Fe] and [O/Fe] values in H4 is in order. The two common methods for measuring [O/Fe] in
red giants and in horizontal branch stars (forbidden [OI] lines at $\sim 6300$ and 6344 \AA\ and the IR triplet) both require
high-resolution spectroscopy. Furthermore, the two methods are at odds with each other, differing by $\sim 0.6$ dex at low metallicities
\citep{C01}. High-resolution spectra and a solution to this discrepancy will be needed before accurate calculations and comparisons of [O/Fe] can be
made.

A qualitative argument for a higher metallicity for H4 can be seen in Figure 1. The metal lines in H4 are clearly stronger than those in H5,
e.g., the Mg2 and Fe5270 index bands. Note in particular the increased strength of the G Band, which measures CH, and is closely correlated with [Fe/H] 
\citep{BH90}. Based \emph{only} on our data, the metallicity difference between H4 and the other clusters is significant at the 2$\sigma$ level.

In sum, we believe that H4 is $\sim 0.3-0.4$ dex more metal-rich than the other Fornax GCs, and that the discrepancy between our spectral [Fe/H] estimate (as
well as those of previous spectroscopic and photometric studies; see Table 5) and that based on the RGB slope could be caused by a solar-like [$\alpha$/Fe]
ratio in the cluster. High resolution spectra are needed to confirm this suggestion.

We offer a final, alternative explanation of the nature of H4 that could resolve the controversy. We find that H4 is old like clusters H1--H3, and yet slightly
more metal-rich. It is also known to be the most compact cluster and is centrally located. We speculate that it is actually the nucleus of the Fornax dSph
galaxy (this same possibility has been suggested by Hardy 2002). As such, it would be comparable to the Galactic GC M54, which has been suggested to be the
remnant nucleus of the Sagittarius (Sgr) dSph galaxy. Like H4, M54 (Layden \& Sarajedini 2000) is as old as Terzan 8 and Arp 2 (two other Sgr GCs) and slightly
more metal-rich (though this result is not statistically significant). Interestingly, Layden \& Sarajedini find evidence for multiple Sgr field star formation
episodes with ages of 11 and 5 Gyr, with the most recent burst occurring 0.5 Gyr ago. As we shall discuss in \S 7, Fornax has a similarly bursty star formation
history.

\section{The Fornax Star Formation History and Discussion}

Buonanno et al.~(1999) found that star formation in Fornax began $\sim 12$ Gyr ago, with bursts occurring $\sim$7, 4 and 2.5 Gyr ago and with the most recent
episode in the last $\sim$0.5 Gyr. The wide field imaging study of 40,000 stars by Saviane et al.~(2000) indicates that the oldest population represents a
relatively small component of the galaxy, with the dominant intermediate population having a mean age of 5.4 $\pm$ 1.7 Gyr. Saviane et al.~also found that
the young burst continued until as recently as 200 Myr ago. This latest star formation episode may be the last, as Fornax appears to have little or no ISM
gas left (Young 1999). 

Recently, Gallert et al.~(2002) obtained spectra of the Ca \sc II \rm triplet feature for individual RGB stars using VLT/FORS. The
resulting age-metallicity relation shows a rapid increase in metallicity from [Fe/H] $\sim$ --2 to --1 at 12 Gyr, followed by a near
constant [Fe/H] $\sim$ --1 until a few Gyr ago.

In terms of spatial distribution, the oldest population is the most extended, indicating that the initial burst occurred over the whole galaxy. Subsequent
bursts were progressively more centrally concentrated, perhaps forming as gas was funneled toward the galaxy center. This indicates an outside-in formation
process extending over several Gyr. It is possible that interactions between Fornax and the Milky Way provided the mechanism for such a process, but
defining the past orbits of these galaxies is difficult (see Piatek et al.~2002 for a recent HST proper motion study of the galaxy).

The GCs H1--H3 and H5 are extremely metal-poor (using previous results for H1) and all can likely be associated with the oldest field star population, despite 
the tentative relative youth of H5. A similar situation is seen with old GCs in other Local Group galaxies (e.g., the Large Magellanic Cloud). H4 has been
discussed earlier---it is centrally located and may be slightly more metal-rich than the other GCs, so it is tempting to associate it with the second burst of
star formation. However, our analysis does not provide evidence that it is younger than the other GCs. If we associate the four metal-poor GCs or all
five of the GCs with the oldest field population, which represents $\la$ 30\% of the galaxy light (Saviane et al.~2000), then the specific
frequency (S$_{N}$) of this population rises from $\sim 20$ to 50--65.

There are several possible interpretations of this result. Assuming that dwarf galaxies are the building blocks of more massive galaxies (as argued in
hierarchical scenarios), then it suggests that early protogalactic fragments may have formed GCs with efficiencies much higher than today. Alternatively, 
Harris (2001) has stressed the interpretation of Fornax and other high S$_{N}$ dwarf galaxies as being \emph{field star poor} rather than GC rich. In one 
possible scenario, winds from massive stars and supernovae associated with the first generation of Fornax stars blow a substantial portion of the remaining 
gas out of the galaxy's shallow potential well, which temporarily halts star formation. The subsequent cooling and infall of part of this gas (which may be 
dynamically connected to interactions with the Milky Way) then forms the basis for later bursts of star formation. 

The present study seems to raise more questions than it answers. Are $\alpha$-element anomalies in H4 responsible for the discrepancy between spectroscopic and
CMD [Fe/H] estimates? If so, what is the cause of these anomalies? A classic scenario would be that H4 formed at the tail end of a starburst, after enrichment
of the interstellar medium by both $\alpha$-element rich Type II supernovae (SNe) and the longer-lived, Fe-bearing Type Ia SNe. This might account for its
higher metallicity, but begs the question as to why it is \emph{older} than the presumably $\alpha$-enhanced GC H5.

We could plausibly conjecture that the Fornax dSph was constructed out of several Searle-Zinn fragments \citep{SZ78}, each of which may have produced one or
more GCs and had its own chemical enrichment period before coalescing into a larger object (see Stetson et al.~1996 for a recent discussion of the varied
Searle-Zinn models).  Such a scenario might then separate out the age and abundance issues discussed above.

Future high resolution spectroscopy of Fornax GC red giants, field stars, and possible candidates for stripped stars will be necessary to untangle
the complex interplay of dynamical and chemical evolution in the galaxy.

\section{Conclusions}

We have analyzed spectra of the five globular clusters of the Fornax dSph. We find that clusters H2, H3, and H5 are very metal-poor with [Fe/H] $\sim -1.8$.
We discuss the controversial case of cluster H4 and conclude that it is slightly more metal-rich than the other four GCs with [Fe/H] $\sim$ --1.5, thus
supporting previous spectral and photometric results but contrary to recent CMD-based claims. We also argue for $\alpha$-element anomalies in H4, similar to
the outer-halo Milky Way GCs Ruprecht 106 and Pal 12. We speculate that H4 may be the nucleus of Fornax, comparable to M54 in the Sagittarius dSph. There is
evidence for an age spread among the five clusters, with H5 possibly younger than the other clusters by a few Gyr.

\newpage
\epsfxsize=14cm
\epsfbox{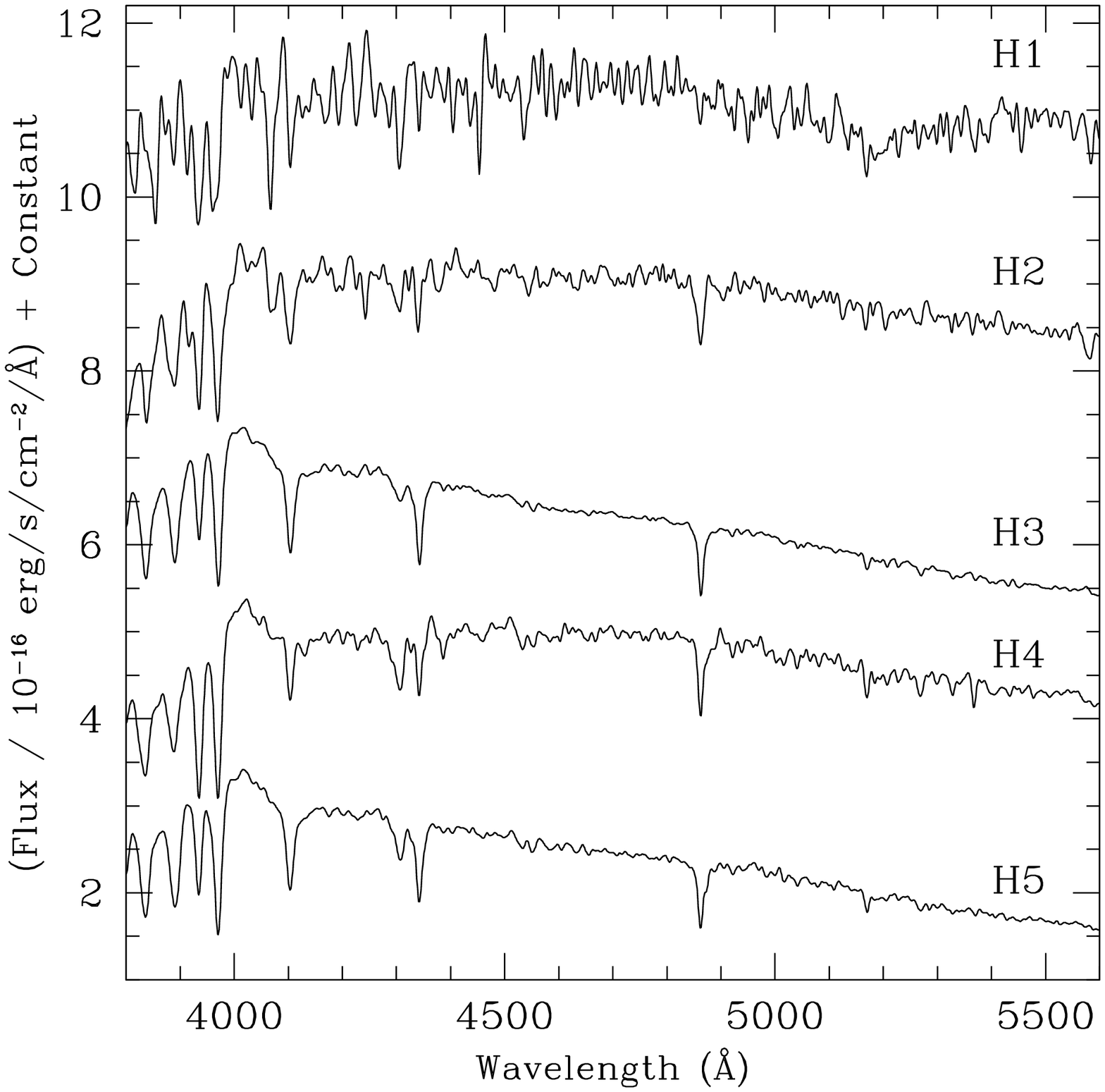}
\figcaption[strader.fig1.ps]{\label{fig:strader.fig1}Flux-calibrated LRIS spectra of Fornax dwarf globular clusters. The spectra have been smoothed to the
Lick/IDS resolution using a wavelength dependent Gaussian kernel. Lick/IDS indices have been measured for clusters H2-H5; the spectrum of H1 has too low S/N
for useful analysis. A constant has been added to the spectra for display purposes.}

\newpage
\epsfxsize=14cm
\epsfbox{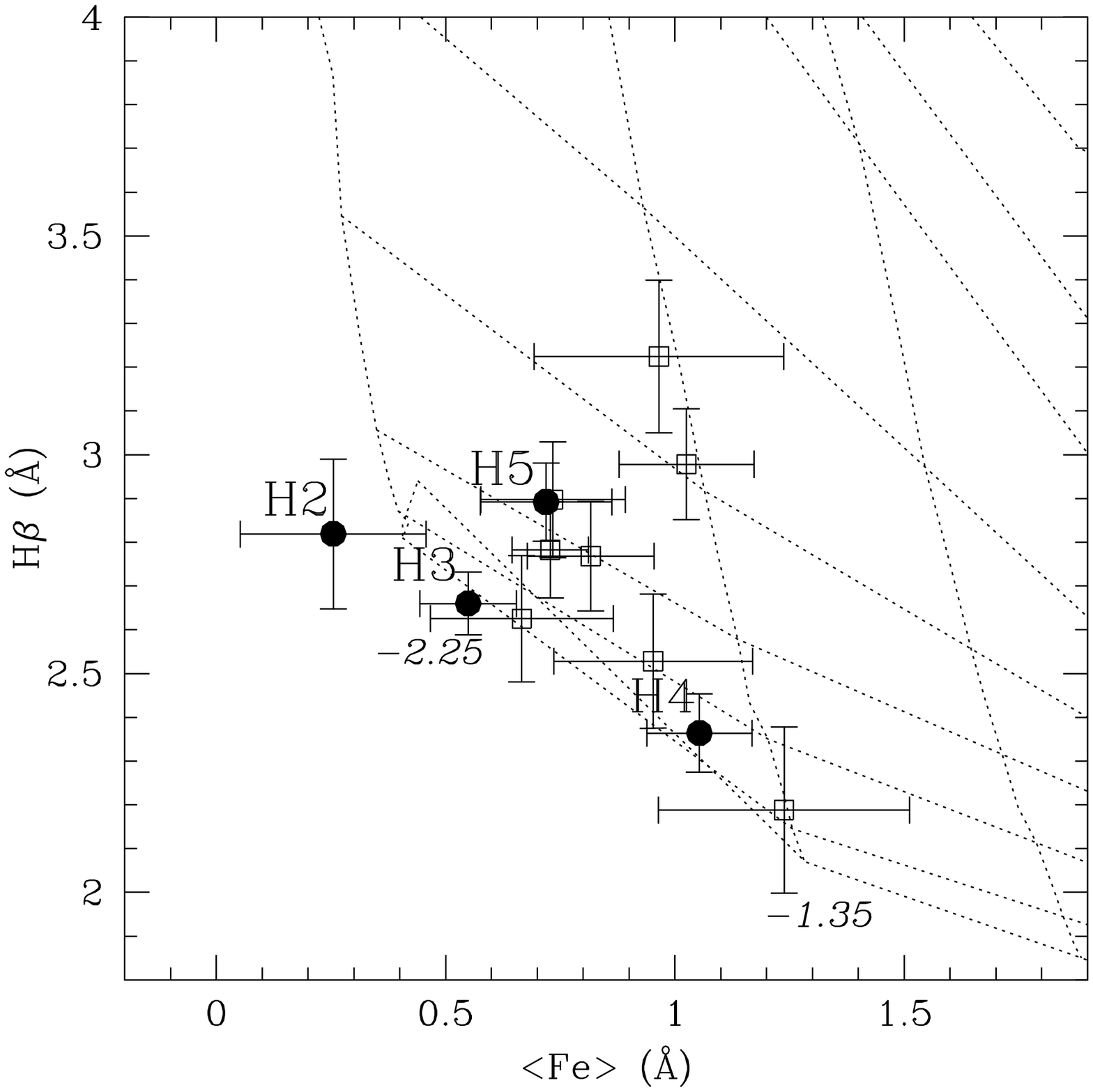}
\figcaption[strader.fig2.ps]{\label{fig:strader.fig2}The Lick/IDS indices $<$Fe$>$ and H$\beta$ for the Fornax dwarf globular
clusters (solid circles) compared to the stellar population models of Thomas et al.~(2002). Identifications for the clusters are given to the upper left
of each point. For comparison, we show the sample of Large Magellanic Cloud globular clusters from Beasley, Hoyle \& Sharples (2002). Numbers in italics denote
model isometallicity lines, and the nearly horizontal lines are isochrones (from bottom to top, 15, 13, 11, 9, 7, 5, 3, 1, and 0.5 Gyr respectively).}

\newpage
\epsfxsize=14cm
\epsfbox{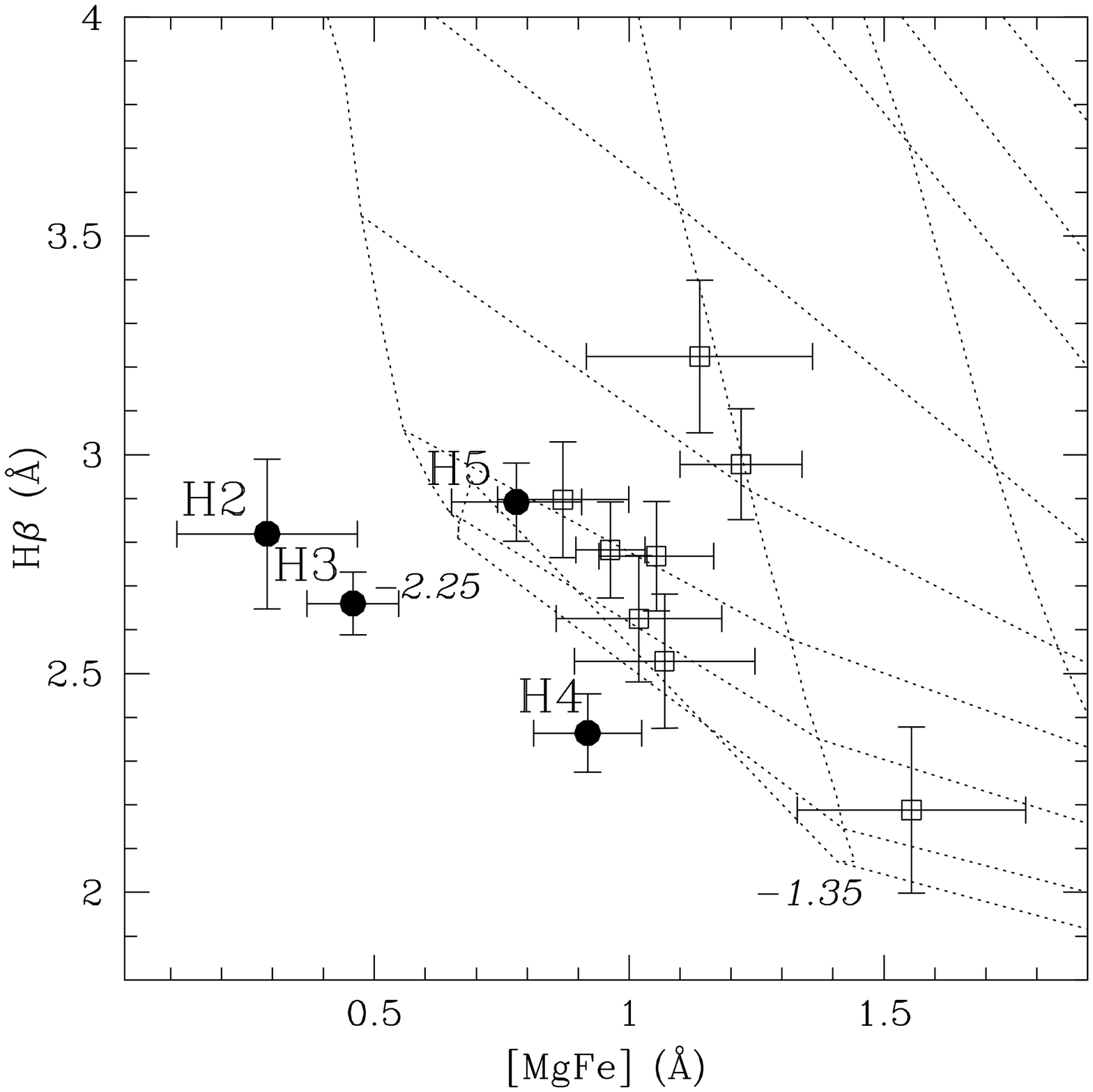}
\figcaption[strader.fig3.ps]{\label{fig:strader.fig3}The Lick/IDS indices [MgFe] and H$\beta$ for the Fornax dwarf globular clusters (solid circles) compared
to the stellar population models of Thomas et al.~(2002). Identifications for the clusters are given to the upper left of each point. For comparison, we show
the sample of Large Magellanic Cloud globular clusters from Beasley, Hoyle \& Sharples (2002). Numbers in italics denote model isometallicity lines, and the
nearly horizontal lines are isochrones (from bottom to top, 15, 13, 11, 9, 7, 5, 3, 1, 0.5 Gyr respectively).}

\begin{deluxetable}{lccccccccc}  
\tablewidth{0pt}
\rotate
\tabletypesize{\scriptsize}
\tablecaption{ \label{tab:indic}
  Lick/IDS Indices}
\tablehead{ID &  CN1   & CN2   & Ca4227 & G4300 & H$\beta$ & $\textrm{Mg}_{2}$ & Mgb   & Fe5270 & Fe5335 \\
              &  (mag) & (mag) & (\AA)  & (\AA) & (\AA)    & (mag)             & (\AA) & (\AA)  & (\AA) }
\startdata

H2 & $-0.069\pm0.008$ & $-0.016\pm0.007$ & $0.07\pm0.11$ & $2.08\pm0.25$ & $2.82\pm0.17$ & $0.011\pm0.004$ & $0.33\pm0.16$ & $-0.16\pm0.15$ & $0.67\pm0.22$ \\
H3 & $-0.073\pm0.003$ & $-0.025\pm0.004$ & $0.11\pm0.06$ & $0.70\pm0.12$ & $2.66\pm0.07$ & $0.011\pm0.002$ & $0.38\pm0.07$ & $0.58\pm0.08$ & $0.51\pm0.11$ \\
H4 & $-0.034\pm0.004$ & $0.003\pm0.005$ & $0.35\pm0.07$ & $2.51\pm0.14$ & $2.36\pm0.09$ & $0.036\pm0.003$ & $0.80\pm0.11$ & $1.14\pm0.11$ & $0.97\pm0.10$ \\
H5 & $-0.076\pm0.005$ & $-0.033\pm0.006$ & $0.16\pm0.08$ & $1.60\pm0.15$ & $2.89\pm0.09$ & $0.025\pm0.003$ & $0.84\pm0.12$ & $0.76\pm0.13$ & $0.68\pm0.13$ \\

\enddata
\end{deluxetable}

\begin{deluxetable}{lccccccc}
\tablewidth{0pt}
\rotate
\tabletypesize{\footnotesize}
\tablecaption{Brodie \& Huchra Metallicity Indices
        \label{tab:metalind}}
\tablehead{ ID  & $\textrm{Mg}_{2}$ & MgH & G Band & CNB & Fe5270 & CNR & H + K}
\startdata

H2 & $0.020\pm0.008$ & $-0.011\pm0.006$ & $0.049\pm0.019$ & $-0.100\pm0.033$ & $-0.006\pm0.009$ & $-0.063\pm0.017$ & $0.437\pm0.034$ \\
H3 & $0.029\pm0.002$ & $-0.014\pm0.002$ & $0.018\pm0.005$ & $0.127\pm0.007$ & $0.014\pm0.003$ & $-0.062\pm0.004$ & $0.149\pm0.005$ \\
H4 & $0.050\pm0.003$ & $-0.007\pm0.003$ & $0.079\pm0.008$ & $0.128\pm0.013$ & $0.028\pm0.004$ & $-0.028\pm0.007$ & $0.227\pm0.010$ \\
H5 & $0.044\pm0.004$ & $-0.016\pm0.003$ & $0.047\pm0.008$ & $0.117\pm0.012$ & $0.018\pm0.004$ & $-0.065\pm0.007$ & $0.182\pm0.009$ \\

\enddata
\end{deluxetable}

\begin{deluxetable}{lcccccccc}
\tablewidth{0pt}
\rotate
\tabletypesize{\footnotesize}
\tablecaption{Metallicity Estimates from Brodie \& Huchra Indices
	\label{tab:metal}}
\tablehead{ ID  & $\textrm{Mg}_{2}$ & MgH & G Band & CNB & Fe5270 & CNR & H + K & [Fe/H]} 
\startdata

H2 & $-2.13\pm0.35$ & $-1.90\pm0.49$ & $-1.88\pm0.38$ & $-2.55\pm0.40$ & $-2.18\pm0.63$ & $-1.75\pm0.47$ & $0.41\pm0.51$ & $-1.76\pm0.41$ \\
H3 & $-2.10\pm0.34$ & $-1.91\pm0.48$ & $-2.22\pm0.32$ & $-1.10\pm0.34$ & $-1.75\pm0.61$ & $-1.75\pm0.45$ & $-2.04\pm0.42$ & $-1.84\pm0.18$ \\
H4 & $-1.84\pm0.34$ & $-1.82\pm0.48$ & $-1.55\pm0.32$ & $-1.08\pm0.34$ & $-1.48\pm0.61$ & $-1.46\pm0.45$ & $-1.40\pm0.43$ & $-1.52\pm0.12$ \\
H5 & $-1.96\pm0.34$ & $-1.93\pm0.48$ & $-1.88\pm0.32$ & $-1.16\pm0.34$ & $-1.65\pm0.61$ & $-1.78\pm0.45$ & $-1.78\pm0.43$ & $-1.73\pm0.13$ \\

\enddata
\end{deluxetable}

\begin{deluxetable}{lcc}
\tablewidth{0pt}
\tablecaption{ \label{tab:indicb}
 Selected Balmer Indices}

\tablehead{ID & H$\gamma_{A}$ & H$\delta_{A}$ \\
              & (\AA) & (\AA) \\}
\startdata

H2 & $1.02\pm0.17$ & $2.54\pm0.26$ \\
H3 & $2.74\pm0.11$ & $3.36\pm0.11$ \\
H4 & $1.27\pm0.15$ & $1.86\pm0.16$ \\
H5 & $2.71\pm0.15$ & $3.37\pm0.17$ \\

\enddata
\end{deluxetable}

\begin{deluxetable}{cccccrr}
\tablewidth{0pt}
\rotate
\tablecaption{Cluster Metallicity Estimates
        \label{tab:compfeh}}
\tablehead{[Fe/H] & [Fe/H] & [Fe/H] & [Fe/H] & [Fe/H] &  & \\
		H1 & H2 & H3 & H4 & H5 & Method & Reference}
\startdata

\nodata & $-1.76\pm0.41$ & $-1.84\pm0.18$ & $-1.52\pm0.12$ & $-1.73\pm0.13$ & Brodie \& Huchra Indices	    & This Work \\
\nodata & $<-2.0$	 & $-2.00\pm0.20$ & $-1.40\pm0.20$ & $-1.85\pm0.25$ & Lick Indices	    & This Work \\
\nodata & \nodata & $-1.93\pm0.20$ & $-1.35\pm0.15$ & $-1.89\pm0.20$ & Mean Line Profile     & Dubath et al. (1992) \\   
\nodata & $-2.00\pm0.78$ & $-1.94\pm0.36$ & \nodata	   & $-2.00\pm0.61$ & Line Indices 	    & Brodie \& Huchra (1991)\tablenotemark{a} \\

\nodata & \nodata &       \nodata  & $-2.01\pm0.20$ & \nodata        & RGB Slope 	    & Buonanno et al. (1999) \\
$-2.20\pm0.20$  & $-1.78\pm0.20$ & $-1.96\pm0.20$ & \nodata        & $-2.20\pm0.20$ & RGB Slope 	    & Buonanno et al. (1998) \\
$-2.0\pm0.4$    & \nodata 	 & $-2.0\pm0.4$   & \nodata        & \nodata 	    & RGB Slope 	    & Jorgensen \& Jimenez (1997) \\
$-2.0\pm0.2$    & \nodata	 & \nodata	  & \nodata	   & \nodata	    & $BV$ photometry       & Demers et al. (1990) \\
\nodata & $-2.1\pm0.2$   & \nodata	  & $-1.9\pm0.2$   & \nodata	    & RGB Slope/Color	    & Beauchamp et al. (1985) \\
$-1.99\pm0.17$  & $-1.73\pm0.17$ & $-2.12\pm0.17$ & \nodata        & $-1.99\pm0.17$ & RGB Slope 	    & Buonanno et al. (1985) \\ 
\nodata & $-1.82\pm0.10$ & $-2.40\pm0.05$ & $-1.45\pm0.03$ & $-1.88\pm0.10$ & Zinn Q	            & Zinn \& Persson (1981) \\
\nodata & $-1.7\pm0.4$ 	 & $-2.1\pm0.4$   & $-1.4\pm0.4$   & $-2.1\pm0.4$   & Washington Photometry & Harris \& Cantera (1977) \\

\enddata

\tablenotetext{a}{The indices themselves are presented in Huchra et al.~(1996).}

\end{deluxetable}

\end{document}